\documentclass{article}

\usepackage{microtype}
\usepackage{graphicx}
\usepackage{subfigure}
\usepackage{booktabs} 

\usepackage{hyperref}

\usepackage[accepted]{icml2023}

\usepackage{amsmath}
\usepackage{amssymb}
\usepackage{mathtools}
\usepackage{amsthm}
\usepackage{bm}

\usepackage{listings}  

\usepackage{hyperref}
\usepackage{url}
\usepackage{multirow}
\usepackage{xcolor}
\usepackage{cleveref}
\usepackage{amsmath}
\usepackage{amssymb}
\usepackage{latexsym}
\usepackage{pifont}
\usepackage{multirow}

\usepackage{mathdots}
\usepackage{todonotes}
\usepackage{tabularx}

\usepackage{subfigure}
\usepackage{booktabs}

\usepackage{algorithm}
\usepackage{algorithmic}
\usepackage{wrapfig}
\usepackage{multirow}
\usepackage[normalem]{ulem}  
\usepackage{mdframed}  

\theoremstyle{plain}

\theoremstyle{definition}

\theoremstyle{remark}

\icmltitlerunning{
Outline, Then Details: Syntactically Guided Coarse-To-Fine Code Generation
}

\def\iodata{I/O data}

\def\dSample{$N_\mathrm{sample}$}

\def\dToken{$N_\mathrm{token}$}

\def\dsample{N_\mathrm{sample}}

\def\dtoken{N_\mathrm{token}}

\definecolor{cc}{rgb}{0.0,0.0,0.0} 

\usepackage{wrapfig}

\begin{document}

\twocolumn[
\icmltitle{
Outline, Then Details: Syntactically Guided Coarse-To-Fine Code Generation
}




\begin{icmlauthorlist}
\icmlauthor{Wenqing Zheng}{yyy}
\icmlauthor{S P Sharan}{yyy}
\icmlauthor{Ajay Kumar Jaiswal}{yyy}
\icmlauthor{Kevin Wang}{yyy}
\icmlauthor{Yihan Xi}{yyy}
\icmlauthor{Dejia Xu}{yyy}
\icmlauthor{Zhangyang Wang}{yyy}
\end{icmlauthorlist}

\icmlaffiliation{yyy}{VITA Group, The University of Texas At Austin, Austin, TX, US}

\icmlcorrespondingauthor{Wenqing Zheng, Zhangyang Wang}{w.zheng@utexas.edu, atlaswang@utexas.edu}

\icmlkeywords{Machine Learning, ICML}

\vskip 0.3in
]



\printAffiliationsAndNotice{}  

\newcommand{\wenqin}[1]{\textbf{\textcolor{orange}{[wenqing: #1]}}}

\begin{abstract}
For a complicated algorithm, its implementation by a human programmer usually starts with outlining a rough control flow followed by iterative enrichments, eventually yielding carefully generated syntactic structures and variables in a hierarchy. However, state-of-the-art large language models generate codes in a single pass, without intermediate warm-ups to reflect the structured thought process of ``outline-then-detail". Inspired by the recent success of chain-of-thought prompting, we propose \mbox{\textbf{ChainCoder}}, a program synthesis language model that generates Python code progressively, i.e. \textit{from coarse to fine} in \textit{multiple passes}. We first decompose source code into layout frame components and accessory components via abstract syntax tree parsing to construct a hierarchical representation. We then reform our prediction target into a multi-pass objective, each pass generates a subsequence, which is concatenated in the hierarchy.
Finally, a tailored transformer architecture is leveraged to jointly encode the natural language descriptions and syntactically aligned I/O data samples. Extensive evaluations show that ChainCoder outperforms state-of-the-arts, demonstrating that our progressive generation eases the reasoning procedure and guides the language model to generate higher-quality solutions. Our codes are available at: \url{https://github.com/VITA-Group/ChainCoder}.
\end{abstract}


\section{Introduction}
\label{sec:intro}

    \begin{figure*}[ht]
        \centering
        \includegraphics[trim={0.7cm 1.2cm 0.7cm 0},clip,width=\textwidth]{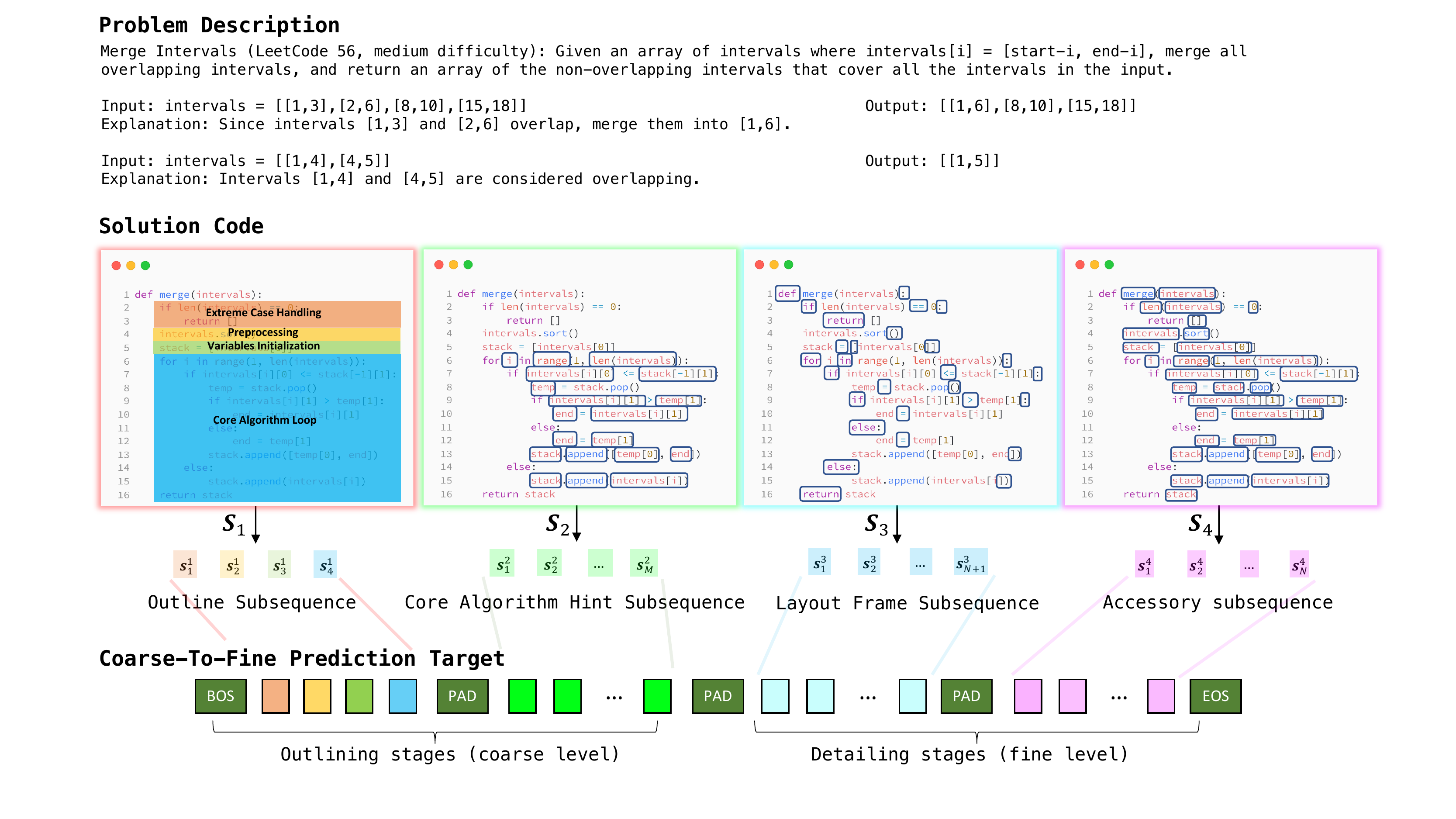}
        \caption{Our prediction target formulation illustrated with an example. In the \textit{Merge Intervals} problem (one medium difficulty problem on leetcode), the solution requires first tackling extreme cases, initializing variables, then entering the main algorithm loop. A programmer needs to first come up with this outline, then think deeper on how to implement the algorithm loop. They may then think about using a stack to keep track of the biggest end time and to decide whether to close or extend the current interval. Finally, after being clear with these ideas, they may write the formal answer with carefulness on both syntax and variable names. This reasoning procedure happens from coarse to fine. We therefore construct the prediction target into four subsequences. Note that the boxed substrings only roughly indicate the tokens, while our actual tokens take the forms of the grouped nodes in the syntax tree. 
        }
        \label{fig:teaser}
    \end{figure*}

    
    The goal of automatic program synthesis has been established and extensively researched for decades \citep{Waldinger1969PROWAS,backus1957fortran}. Recently, with the rapid advancements of Large Language Models (LLMs), a surge of methods leverage transformers to model programs as sequences, and have demonstrated prospects to automatically analyze, annotate, translate, or synthesize code \citep{li2022competition, austin2021program, fried2022incoder, chen2021evaluating, Kanade2020LearningAE, feng2020codebert, Clement2020PyMT5MT, Svyatkovskiy2020IntelliCodeCC, Wang2021CodeT5IU}.
    These LLMs are trained on a large corpus of code so as to imbibe syntactic and semantic understanding into model weights, and develop logical reasoning for program synthesis.

    Though current LLMs have gained success in code understanding and generation, their inherent design, as well as the training/finetuning techniques, are largely borrowed from the \textit{natural language} modeling. This limits their potential in code generation modeling due to two reasons. 
    
    First, for a given problem specification, LLMs generate end-to-end code solutions autoregressively in a single pass regardless of the logical complexity involved. Even from a programmer's perspective, it is difficult to write good code in a single shot without laying out logical and hierarchical thinking.
    In other words, composing code involves molding an overall ``warm-up'' proceeded by lower-level atomic algorithms and variable definitions\footnote{This is usually termed as a \textit{top-down practice} in programming: beginning with creating an outline or high-level view of the program, and then successively breaking it down into smaller and smaller components until the entire program is complete.}. 
    
    Second, LLMs tokenize the code using tokenizers meant for everyday-natural-language strings and disregard syntactical awareness as structural priors \citep{chen2022codet,hendrycks2021measuring,li2022competition,chen2021evaluating,austin2021program,athiwaratkun2022multi,nijkamp2022codegen,xu2022ide,wang2022code4struct,chensparse,li2023cancergpt,mao2022single}.
    Notably, programming languages contain complex rules and conform to more rigorous layouts compared to their natural language counterparts \citep{casalnuovo2019studying, naur1975programming}. They have fewer choices of synonyms, stringent syntax requirements, and diverse control flows (sequential statement execution, selection/branching, and repetition such as for and while loops). 

    In our ChainCoder, we investigate if addressing the aforementioned flaws is indeed promising. As shown in \Cref{fig:teaser}, we aim to remodel the code generation that previous works treat as a no-warm-up, single-shot textual sequence completion task into a multi-step intermediate reasoning strategy through syntactical decomposition. 

    
    In a similar context of harnessing multi-step reasoning capability of LLMs (on natural language), recent works discover that chain-of-thought (CoT) prompting \citep{zhang2022automatic, wei2022chain,  shi2022language} significantly boosts performance. CoT prompting advances LLMs to naturally develop a series of intermediate reasoning steps before providing the final answer through exemplar demonstrations. This exposes a profound potential of refactoring LLMs to explicitly perform \textbf{coherent intermediate step-wise reasoning}, rather than directly concluding final solutions. 


    Inspired by the success of chain-of-thought prompting for LLMs, we design a novel approach
    that follows a \textit{step-wise}, \textbf{outline-then-detail} thought process. 
    We decompose the source code into a predefined hierarchy of code logic, and prepare the LLM generation target in each level ($\bm{S}_{1-4}$ in \Cref{fig:teaser}) so that the model predicts them in multiple passes separately.
    We refer to our model as ChainCoder, and such hierarchical prediction as ``\textit{coarse-to-fine}'' generation, similar to \citet{dong2018coarse}.
    Specifically, ChainCoder first builds a logical skeleton of the code, then move to step-wise implementation of lower-level atomic details/algorithms. This is inherently different from vanilla CoT prompting, as CoT methods require external prompting and are mainly developed in the context of natural language reasoning, while ChainCoder is the first program synthesis model that progressively generates code by itself.

    To build such a multi-level hierarchical representation, appropriate disentanglement of the code is required.
    To this end, we tailor the tokenization step to better display complex programming language rules. 
    We parse the code into the Abstract Syntax Tree (AST), which naturally exposes extensive domain knowledge and structures expressed in the tree nodes and edges.
    We then disentangle the syntax tree into two components: the \textit{layout frame component}, which tells more clause type and other syntactical information, and the \textit{accessory component}, which includes concrete variable and function names, etc. 
    The syntax tree-based tokenizer brings advantages in two-fold: its domain knowledge eases the comprehension of source code, and would possibly boost the syntactical correctness. 
    Unlike previous research on applying syntax aware tokenizer \citep{jimenez2018impact} or simplification tricks during the tokenization step \citep{gupta2017deepfix,chirkova2020simple}, ChainCoder is the first approach to leverage syntax knowledge to disentangle the code and build hierarchical prediction targets.

    

    In brief, our work aims to reformulate program synthesis from a sequential text completion task into a multi-pass coarse-to-fine refinement strategy. As depicted in \Cref{fig:teaser}, this comprises of four progressive steps -- outlining, core algorithm hinting, layout framing, and accessory sub-sequencing inspired from the ``warm-up'' thought process of programmers.
    We then leverage a tailored transformer architecture that better encodes the syntactically aligned \iodata{} to ease comprehension of problem specifications. To the best of our knowledge, ChainCoder is the first large language model to perform multi-pass \textit{coarse-to-fine} code generation to reflect intermediate warm-up procedure, in contrary to existing single-shot sequential completion works. In summary, our technical contributions are as follows:
    \begin{itemize}
        \item Inspired by the chain-of-thought process, we leverage the unique syntax hierarchy to build the multi-pass coarse-to-fine code generation framework. ChainCoder is the first program synthesis approach that harnesses progressive generation to elicit the step-wise reasoning capability of LLMs and improve accuracy and syntactic coherency.
        \item To enable progressive generation, we propose a syntax-aware tokenizer that neatly disentangles the code and outputs multi-level prediction targets accordingly. We further develop a transformer to better leverage the structure of the syntactically aligned data.
        \item Evaluations on the competition-level datasets show that ChainCoder performs better than other state-of-the-art models even with smaller model sizes. Ablation studies verified the effectiveness of the coarse-to-fine guidance and other design choices.
    \end{itemize}

\section{Related Works}
    
    

    Synthesizing program from the description and I/O pairs \citep{Chen2018TowardsSC,bunel2018leveraging,devlin2017robustfill,gulwani2012spreadsheet,fox2018parametrized,Ganin2018SynthesizingPF,chen2019execution,hong2021latent,le2022coderl} or other modality inputs \citep{sun2018neural,liu2019learning,tian2019learning} is a well-received benchmark, yet is challenging due to the indefinite program space.

    The classical approaches to program synthesis date back to \textit{rule-based program synthesis} which use formal grammar to derive programs from well-defined specifications \citep{Waldinger1969PROWAS, Manna1971TowardAP, Manna1980ADA}. Later on, symbolic and neuro-symbolic techniques \citep{balog2017deepcoder,odena2019learning,Ellis2018LearningLO,Ellis2020DreamCoderGG,devlin2017robustfill,Panchekha2015AutomaticallyIA} have also been explored to generate code with higher quality. However, these methods are mainly applied to domain-specific languages (DSLs), which limits the applicability in more advanced programming languages.


    \textbf{Program Synthesis With LLMs.}
    Recently, there has been a huge surge in exploiting language models for program synthesis with extension to general-purpose programming languages \citep{austin2021program,hendrycks2021measuring,chen2021latent,Clement2020PyMT5MT,Wang2021CodeT5IU}. 
    These models are trained on massive codes and natural language datasets, and learn to condition the code on the natural language descriptions or existing code fragments.
    For example, \citet{devlin2017robustfill} uses an encoder-decoder network formulating synthesis process as a sequence generation problem.
    CodeT5 \citep{Wang2021CodeT5IU} 
    prominently focuses on understanding tasks such as code defect detection, translation, and clone detection. Codex \cite{chen2021latent} uses GPT-3 architecture, evaluating its synthesis performance on a new benchmark of simple programming problems. CodeBERT \cite{feng2020codebert} is pre-trained for natural language and programming languages like Python, Java, JavaScript, etc., and captures the semantic connection between natural language and programming language. 

    More recent state-of-the-art works such as APPS \citep{hendrycks2021measuring} and AlphaCode \citep{li2022competition} have shown promising results over competition-level problems.
    Another related program synthesis framework InCoder \citep{fried2022incoder} is able to refine the code via infilling under a bidirectional context.
    Similarly, Parsel \citep{zelikman2022parsel} generates programs by combining hierarchical generations of sub-programs during interactions between the LLM and an external environment. 
    However, these methods still adopt natural language tokenizers that overlook the special syntax structures of the programs and do not execute the necessary warm-up before generation.

    \textbf{Multi-Step Reasoning In Language Modeling.}
    Recent years have witnessed a shift of LLM paradigm from ``pre-train, fine-tune'' procedure into ``pre-train, prompt, and predict'' \citep{liu2021pre}. Among the prompting methods, the chain-of-thought (CoT) has become the research highlight \citep{zhang2022automatic, wei2022chain,  shi2022language}.
    Chain of thought refers to techniques that provide intermediate reasoning steps as prompting demonstrations for language models \citep{zhang2022automatic}.
    
    The CoT methods allow LLMs to decompose problems into intermediate steps. They provide an interpretable window into the model behaviors to ease the debugging, and can be readily elicited by adding examples in the few-shot prompting \citep{wei2022chain}. 
    They have also been demonstrated to improve model performance in-situ for multiple choice question answering tasks \citep{lewkowycz2022solving}, and to better perform in multi-step reasoning task such as the math word problems \citep{chowdhery2022palm}.
    
    \textbf{Generating Programs Hierarchically From Abstractions}
    Recently, a few papers aims to improve the scalability by generating certain forms of sketches first, then convert to full operational programs. Specifically, 
    \citet{murali2017neural} starts with generating sketches that leave out names and operations. They then infer a posterior distribution over sketches, from which to samples type-safe programs using combinatorial techniques.
\citet{nye2019learning} follows previous strategies to break code generation into two steps: to generate sketches and to fill full in them. Previous methods assign the generation step with pattern recognition or perform symbolic search \citep{zheng2022symbolic,zheng2022symbolic1} in a fixed way. This paper proposed a novel algorithm that learns when to switch between the two ways without direct supervision.
\citet{brockschmidt2018generative} uses a graph to represent the intermediate state of the generated output. It generates code by interleaving grammar-driven expansion steps with graph augmentation and neural message-passing steps. Similarly, \citet{zhong2023hierarchical} separately learned two stages: (1) a high-level module to comprehend the task specification from long programs into task embeddings, and (2) a program decoder that can decode a task embedding into a program.

In comparison, ChainCoder uses a unified space to represent the middle step sketches and the final representations, instead of specifically curating different forms of representations. On the other hand, the entire decoding mechanism of ChainCoder is learned by a single LLM, and the results are also directly sampled from this LLM. The decomposition of the program into syntax and components is explicitly done via a novel tokenizer, also in the form of a unified string.


\section{Coarse-To-Fine Programming Language Modeling With ChainCoder}
    \label{sec:method}
    
    
    
    This section discusses how ChainCoder tokenizes and predicts code in a hierarchical fashion. In brief, the ChainCoder first lays out a coarse-level outline of the code, then focuses on fine-grained details. The coarse-level layout can be thought of as the summarized representation of constituent fine-level parts. In order to better leverage code structure and ease learning, our tokenization does not base on code as raw text but rather as a specialized data structure parsed from source code -- namely, the abstract syntax tree (AST).
    In the following, we first discuss the preliminaries for AST parsing in \Cref{section:ast}, and formulate the tokenization algorithm in \Cref{sec:encoding-alg}. We then discuss the construction of multi-step prediction targets in \Cref{sec:prediction-target} and finally describe ChainCoder's model architecture in \Cref{sec:model arch}.
    
    
\subsection{Preliminaries of Parsing and Abstract Syntax Tree}
    \label{section:ast}
        Codes are written by humans, but are executed by computers. Therefore, humans write them in a form that they can understand, then the software transforms them in a way that can be used by the computer.
        Such transformation is referred to as parsing, which determines a structured template from the source code text.
        After parsing, the programming language is converted to a better-formatted data structure, usually a tree-type format called the abstract syntax tree. 
        
        Parsing the code into the tree-type format is common to almost all programming languages. Specifically in Python, the leaf nodes of the abstract syntax tree are the function/class/variable/operator names and constant values. The branching nodes of the tree represent syntactic relationships between the leaf nodes. 
        For example, consider the line of code \texttt{x = 0}. Its syntax tree is roughly a tree with three nodes, with the root node being the \textit{assignment}, and two leaf nodes being \texttt{x} and \texttt{0}. The root node conveys syntax information (``$\mathrm{=}$''), which tells that this line is an ``assignment'' type sequential statement, there will be two placeholders as its children nodes, and one child node will assign its value to the other child node. The precise syntax tree is more complex though, which we show in \Cref{fig:SMALLAST} and \Cref{fig:LARGEAST} in the Appendix.
        The abstract syntax tree has better-formatted structures than the raw text, but is overlooked by previous LLM research on program synthesis. Therefore, we seek to ease the learning of the LLM by tokenizing based on the parsed tree rather than raw code text.

\subsection{Syntax-Aware Tokenization: Layout Frame and Accessory Components}
\label{sec:encoding-alg}
The code contains complex structures both syntactically and semantically. Writing algorithmic code is generally difficult, partially due to the fact that both syntactic and semantic information need to be handled simultaneously. To ease this difficulty, we propose to disentangle the code into a component that conveys more syntax information and another component that conveys more semantic information and generate them separately.
We refer to these two components as the \textit{layout frame subsequence} and the \textit{accessory subsequence}.
The disentanglement procedure is based on the syntax tree representation, as it enables more flexible treatments of the code in the domain of tree nodes and edges.

The layout frame subsequence is responsible for indicating the clause type, assigning placeholders in the sequel, and maintaining syntactic correctness. On the other hand, the accessory subsequence contains the concrete variable/class/function/operator names, and constant values.
In the above example of \texttt{x = 0}, the syntax subtree of this line of code could be unambiguously decoupled into two layout frame tokens and two accessory tokens. 
First, the two accessory tokens are easily identified as the leaf nodes, \texttt{x} and \texttt{0}. Then, the first layout frame token could be the root node \textit{assignment (=)} with the corresponding edge between it and the leaf node \texttt{x}, while the second layout frame token could be the edge between the root node and the other leaf node \texttt{0}. 
When these tokens are interleaved together, they become the serialized representation of the syntax tree.

The training procedure of ChainCoder heavily relies on an \textit{encoding} step, which converts the source code into decoupled layout frame and accessory subsequences. 
In the encoding step, the code is first parsed into an AST, which is then partitioned into nodes and edges. These nodes and edges are then re-grouped to generate these two subsequences. The resulting subsequences are used for training and inference. 
On the other hand, we use the term \textit{decoding} to refer to the reverse procedure. Given the decoupled subsequences, the decoding procedure recovers the original source code. It is noteworthy that the encoded two subsequences will be less human-interpretable than the original code, yet they convey precise information to the Python interpreter.

As revealed in \Cref{fig:teaser}, the layout frame and the accessory subsequences constitute the fine level part in the coarse-to-fine hierarchy, and are referred to as $\bm{S}_3$ and $\bm{S}_4$.
They contain complete information to reconstruct the original code by themselves. There are other parts of coarse level subsequences ($\bm{S}_1$ and $\bm{S}_2$), which we discuss in \Cref{sec:prediction-target}. Next we discuss the encoding and decoding steps in detail.

\textbf{Encoding Step.}
The encoding step is used to preprocess the training samples to construct the prediction targets. Prior to training, each code sample in the training set is converted to the coarse-to-fine hierarchical subsequences, and the ChainCoder is trained to generate these new subsequences.

The encoding step first parses the source code into the abstract syntax tree, then disentangles the serialized syntax tree into the layout frame subsequence ($\bm{S}_3$) and the accessory subsequence ($\bm{S}_4$). It constructs the other two coarse-level subsequences at the same time.
The encoding algorithm of ChainCoder is displayed in \Cref{alg:wordy}.


In general, the encoding step is completed during the depth-first pre-order traversal for the parsed syntax tree. Starting from the root node, it first visits the current node, then visits all the children nodes from left to right. Every time it meets a leaf node, the traversal pauses, and it groups all branching nodes seen so far into a token, and appends the grouped branching nodes to the layout frame subsequence. It then appends the leaf node into the accessory subsequence. 
By grouping the branching nodes, the algorithm efficiently compresses token sequence length.
The following relationship always satisfies for the tokenized subsequences: \texttt{layout frame subsequence length = accessory subsequence length + 1}, where the ``\texttt{+1}'' comes from the ending brackets in the layout frame subsequence after the last accessory token is appended.

\begin{algorithm}[H]
\caption{The Encoding Step Of ChainCoder}\label{alg:wordy}
\begin{algorithmic}
\REQUIRE{Source code}
\ENSURE{Four subsequences: outline $\bm{S}_1$, core algorithm hint $\bm{S}_2$, layout frame $\bm{S}_3$, accessory $\bm{S}_4$}
\end{algorithmic}
\begin{algorithmic}[1]
\STATE Parse the source code into syntax tree $\phi$
\STATE Initialize $\bm{S}_1, \bm{S}_2, \bm{S}_3, \bm{S}_4$ with empty sequences
\STATE Branching nodes group $t \gets \mathrm{empty\ string}$
 \WHILE{Traversal not finished}
  \STATE Get next node $\bm{\xi}$ using pre-order traversal for $\phi$
  \IF{$\bm{\xi}$ is branching node}
        \STATE Group with previous branching nodes: $t \gets t+\bm{\xi}$

  \ELSIF{$\bm{\xi}$ is leaf node}
        \STATE Append grouped branching nodes: $\bm{S}_3 \gets \bm{S}_3 + t$
        \STATE Append leaf node: $\bm{S}_4 \gets \bm{S}_4 + \bm{\xi}$
        \IF{$\bm{\xi}$ is within a loop or user-defined function}
        \STATE Append packed branching nodes: $\bm{S}_2 \gets \bm{S}_2 + t$
        \ENDIF
        \IF{$\bm{\xi}$ is the first leaf node in the current line and line indent level is 1}
            \STATE Add branching nodes to outline: $\bm{S}_1 \gets \bm{S}_1 + t$
        \ENDIF
        \STATE $t \gets \mathrm{empty\ string}$
  
  \ELSIF{$\bm{\xi}$ is the starting token of a new line}
        \STATE Remove the doc-strings and comments, if any
        \FOR{All variable names $v_i$ in this line}
            \IF{$v_i$ is not linked to any imported name, and is not a built-in name}
            \STATE Replace $v_i$ in $\bm{S}_4$ with a name pool candidate
            \ENDIF
        \ENDFOR
  \ENDIF

  \ENDWHILE
\end{algorithmic}
\end{algorithm}

The encoding algorithm offers two optional ways to further simplify the generation task: the \textit{name replacement} and the \textit{doc-strings/comments removal}. The name replacement feature detects all user-defined names in the parsed tree, and replaces them from a name pool. In this way, the code logic is maintained, while the extra learning burden caused by naming style differences can be avoided.
The doc-strings/comments removal also naturally takes advantage of tree-based representation, they reduce the sequence length without hurting the functionality.


By parsing and encoding, the program space is reduced and the learning task is simplified. On one hand, thanks to the degenerated variable names, the vocabulary could be shrinked while still leading to clearer meanings. For example, if there are two variables called \texttt{student\_num} and \texttt{student\_nums} in one program, ChainCoder might tokenize them into two distinct words \texttt{user\_defined\_var\_7} and \texttt{user\_defined\_var\_8} drawn from the name pool, rather than using the original names that might cause confusion. The name pool size is also much smaller than the natural language vocabulary which includes all possible sub-words of variable names. On the other hand, the token sequence could be shortened by removing the doc-strings and comments without hurting the functionality.
Some sentences that would be tokenized into long sequences by natural language tokenizer can now be represented very concisely in the syntax tree domain, such as \texttt{for i in range(num)}, \texttt{res = a if x is None else b}. 
What's more, the names of variables/functions/classes/etc are never broken into multiple sub-words in the syntax tree representation.

\textbf{Decoding Step.}
The decoding step is used during the inference time. As ChainCoder generates the hierarchical subsequences rather than the source code, this step is applied to convert the resulting subsequences back to the human-readable source code.

Given the design of the encoded subsequences, the decoding algorithm is straightforward. We first interleave the tokens in $\bm{S}_3$ and $\bm{S}_4$ and glue them together, so that the serialized syntax tree is obtained. Then, an abstract syntax tree unparse tool is used to convert the syntax tree into the source code.

\begin{figure*}[t]
        \centering
        \includegraphics[width=0.8\textwidth]{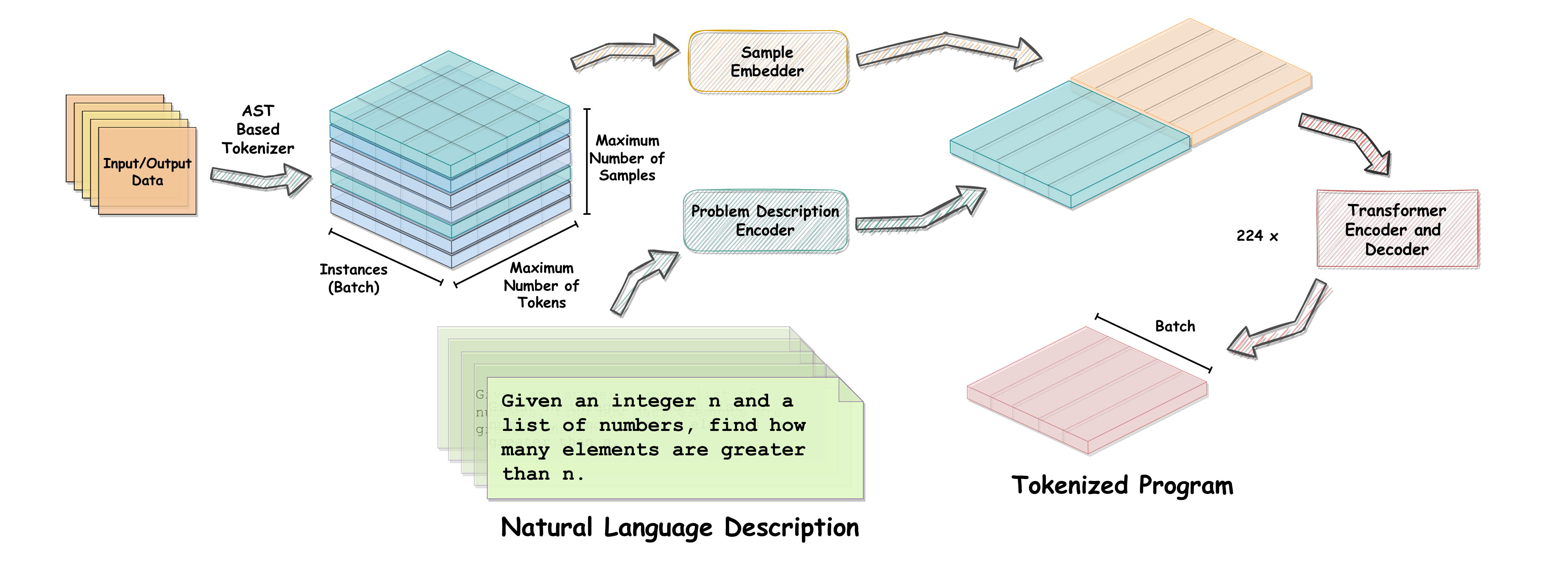}
        \caption{Overall Model Architecture of ChainCoder, which consists of the sample embedder, the problem description embedder, the traditional transformer encoder, and decoder.}
        \vspace{-0.5em}
        \label{fig:model arch all}
\end{figure*}

\subsection{Multi-Level Coarse-To-Fine Prediction Targets}
\label{sec:prediction-target}

        As visualized in \Cref{fig:teaser}, the prediction target sequence is composed of four components: the outline ($\bm{S}_1$), the core algorithm structure hint ($\bm{S}_2$), the layout frame ($\bm{S}_3$) and the accessory ($\bm{S}_4$) subsequences. 
        These four subsequences are concatenated by the system-level \texttt{<PAD>} token.
        
        In the coarse outline $\bm{S}_1$, each token essentially corresponds to a line of code with the smallest indent. 
        For each of these lines, one layout frame token is selected, which is the first token, the one that contains the root node of the subtree of this line.
        The $\bm{S}_1$ is a discontinuous subset of the entire layout frame subsequence $\bm{S}_3$.

        The core algorithm structure hint $\bm{S}_2$ is designed to warm-up the model before generating the final solution ($\bm{S}_3$ and $\bm{S}_4$).
        Since most algorithms contain loops, and the core functional parts are usually implemented in the loops or user-defined functions, we identify the implementations within the loop or user-defined functions as the core algorithms and use their layout frame tokens to construct $\bm{S}_2$.
        This part is a continuous subset of $\bm{S}_4$.
        
        The construction of the last two subsequences $\bm{S}_3$ and $\bm{S}_4$ has been discussed in \Cref{sec:encoding-alg}.
        With these subsequences, ChainCoder breaks the code generation task from a single pass generation into four turns. In each turn, ChainCoder focuses on a simpler subtask, i.e., only generates a specific subsequence. 
        
        The functionalities of $\bm{S}_1$ and $\bm{S}_2$ are analogous to the chain-of-thought prompting \citep{zhang2022automatic,wei2022chain,shi2022language}, except that they are generated by the model itself rather than external prompting. 
        They only exist to hint the model generation procedure. When the entire solution has been generated, they will be removed before checking the correctness of this answer.
        Such design follows the simple philosophy: an outline of the algorithm shall be built before writing down the concrete answer, and a layout frame shall be constructed before the detailed variable names / constants / etc are filled in.

        We adopt the same autoregressive generation recipe as other LLMs, which asks ChainCoder to predict the next token given all previously generated tokens. The inference procedure is also the same: ChainCoder completes the target sequence token by token until a \texttt{<EOS>} token is generated. 
        As $\bm{S}_1$ and $\bm{S}_2$ need not to be complete, we apply token dropout to let the model occasionally generate $\bm{S}_3$ and $\bm{S}_4$ with information in $\bm{S}_1$ and $\bm{S}_2$ missed out. We achieve this by randomly dropping tokens in $\bm{S}_1$ and $\bm{S}_2$ with probabilities 0.05 and 0.2, respectively.

\subsection{The Model Architecture of ChainCoder}
\label{sec:model arch}
        
The inputs of program synthesis contain the problem description in natural language and a group of \iodata{}.
Thanks to the syntax tree parsing, the \iodata{} could be well-aligned across samples based on the values' syntax roles. 
After alignment and padding, the tokens of \iodata{} are arranged into a regular matrix with both sample and token dimensions. 
The natural language descriptions can be well handled by LLMs such as BERT. 
However, the one-dimensional sequence modeling of LLMs fails to fully leverage the structures of the aligned \iodata{}.
We accommodate this gap with a new transformer architecture, which has a sample embedder submodule that attends to the regular matrix-shaped \iodata{}.
The overall model architecture consists of four components: a sample embedder, a description embedder, a token encoder, and a program decoder. The architecture is shown in \Cref{fig:model arch all}.

\def\numinst{N_\mathrm{inst}}
\def\nopt{L_\mathrm{token}}
\def\bR{\mathbb{R}}

\textbf{Sample Embedder.}
With the syntax-based encoding, alignment, and padding, the \iodata{} in each instance is now represented as a matrix $W\in \bR^{\dsample \times \dtoken\times E}$, where \dToken\ is the maximum number of tokens across different \iodata{} samples, \dSample\ is the number of samples for this instance, and $E$ represents the token embedding dimension. 
The sample embedder processes the matrix-shaped aligned \iodata.
To track the locations of tokens in the matrix, the tokens are added with two sets of positional embeddings along the token and sample dimensions.
At the output side of the sample embedder, it uses first element pooling to reduce the embedding size of each instance to $\dtoken \times E$.

\textbf{Description Embedder.}
Each instance in the \iodata\ corresponds to a natural language problem description. We utilize the BERT model \citep{devlin2018bert} to perform natural language understanding, and distill its outputs into four tokens: the first token, the last token, the minimum pooling token, and the maximum pooling token. The output dimensions for one instance is $4 \times E$.

\textbf{Token Encoder, Program Decoder And Training Loss.} 
From the sample embedder and the description embedder, we get the embeddings for the \iodata{} and the problem descriptions separately. 
We concatenate these embeddings and get a sequence with dimensions $(\dtoken + 4) \times E$.
We then follow the traditional transformer encoder and decoder layers and training recipes. At the output of the decoder, we ask the model to predict the coarse-to-fine subsequences described in \Cref{sec:prediction-target} using cross-entropy loss.

\section{Experimental Results}

In our work, we target improving the performance of language models on code generation tasks particularly on competition-level problem-solving. Following previous works in this domain \citep{hendrycks2021measuring, li2022competition, chen2021evaluating, austin2021program}, we leverage the CodeParrot GitHub-Code \citep{codeparrotgithub} dataset for model-pretraining on general purpose source code. We then fine-tune on the training sets of competitive programming benchmarks and evaluate on their test sets respectively. To prevent the data leakage, we execute extra carefulness and strictly followed the datasets as well as the processing steps of AlphaCode and CodeRL \citep{le2022coderl}.

\begin{table}[ht]
    \centering
    \resizebox{0.48\textwidth}{!}{
        \begin{tabular}{l||ccc|ccc}
            \toprule[1.5pt]
            \multicolumn{1}{c||}{Sequence}  & \multicolumn{3}{c|}{APPS \citep{hendrycks2021measuring}}  & \multicolumn{3}{c}{Contests \citep{li2022competition}} \\
            \cmidrule{2-7}
            \multicolumn{1}{c||}{length} & ChainCoder & BERT & Raw String & ChainCoder    & BERT    & Raw String   \\
            \midrule
            Mean &  $155.808$ & $176.312$ & $416.419$ & $145.676$ & $176.509$ & $432.566$ \\
            Median & $113.0$ & $124.0$ & $278.0$ & $111.0$ & $124.0$ & $289.0$ \\
            Std. & $272.911$ & $261.470$ & $618.100$ & $82.547$ & $292.139$ & $780.052$ \\
            \bottomrule[1.5pt]
        \end{tabular}
    }
    \caption{The tokenization sequence length statistics. The syntax-aware tokenizer sequence length is calculated by the summation of layout frame and accessory subsequences. The ChainCoder tokenizer leads to shorder sequence thanks to the domain knowledge built-in to syntax tree.}
    \label{tab:stats}
\end{table}

\textbf{Datasets and Experimental Overview.}
The CodeParrot GitHub-Code pre-training dataset contains $7.2$ million Python files with a total size of $52.03$ GB. These pre-training codes do not contain corresponding \iodata{} pairs, and many also lack natural language descriptions of their functionalities. 
In order to approach the absence of \iodata{}, our sample-embedder submodule output is initially neglected and is only updated during the fine-tuning phase.
Meanwhile, to combat the lack of natural language descriptions, we employ a pre-trained CodeT5-based \citep{Wang2021CodeT5IU} Python code explanation model to generate natural language descriptions for each code sample. The natural-language-embedder-submodule is a pre-trained BERT \citep{devlin2018bert} and is fixed during the pre-training phase, and optimized during fine-tuning.

For the fine-tuning and evaluation, we choose two carefully curated datasets, CodeContests \citep{li2022competition} and APPS \citep{hendrycks2021measuring}, which are both massive and contain problems of varied difficulties (see \Cref{tab:results}). The CodeContests dataset comprises of data scraped from Codeforces along with existing data from Description2Code \citep{Caballero_Description2Code_Dataset_2016}. Meanwhile, APPS is scraped from of Codewars, AtCoder, Kattis, and Codeforces. After pre-trained on the GitHub codes, we use two copies of the model to fine-tune on the training sets of CodeContests and APPS, and evaluate on their test sets respectively.

\begin{figure}[h]
        \centering
        \includegraphics[width=\linewidth]{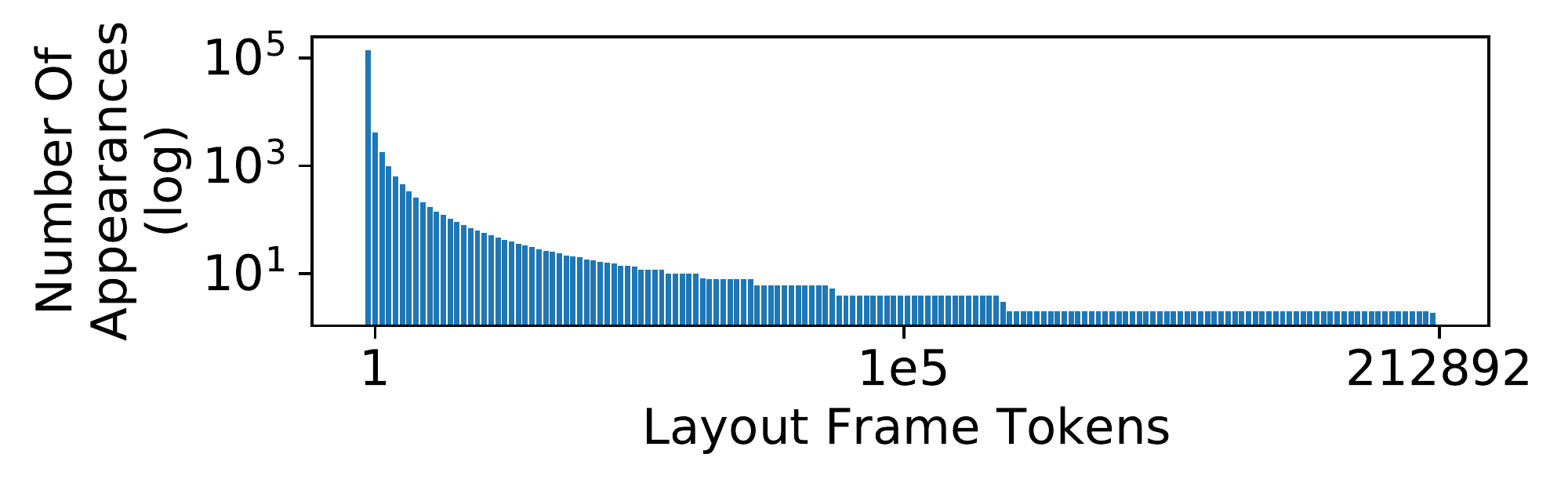}
        \vspace{-2em}
        \caption{Token frequency distribution. Visibly, the majority of tokens are rarely met. From our inspection, these highly infrequent tokens are not necessarily tied to the code solutions but rather just happened to occur randomly. Therefore, it is safe to assume that ChainCoder can be trained on the more-frequent end of the distribution alone without facing any performance losses. 
        }
        \label{fig:vocab-dist}
\end{figure}

\textbf{Tokenization.} 
Before pre-training, we sweep across the APPS and CodeContests datasets to collect \textit{vocabulary}, yielding $212,892$ distinct layout frame tokens. These tokens follow a long tail distribution as illustrated in \Cref{fig:vocab-dist}. Our tokenization step filters out some code files in the dataset that yield too long or incompatible sequences. After our tokenization step, we obtain $81,339$ code files from APPS, and $1,337,655$ code files from CodeContests. 

\textbf{Architectural Details and Metrics.}
ChainCoder contains two sample-embedder transformer blocks, two transformer encoder blocks, 224 transformer decoder blocks, with the 512 hidden dimensions for these submodules, yielding 1.09 billion parameters in total. The natural-language-embedder inherits weights of the pre-trained BERT model, while all other submodules of ChainCoder are trained from scratch. We diversify the learning difficulty during the fine-tuning phase by periodically changing the number of \iodata{} pairs (sweeping between 1 and 32, with a 15 epoch period) and the number of programs that the model predicts (sweeping between 1 and 8, with a 37 epoch period).
At inference time, we set our beam-search width to 5. We adopt the same evaluation metric as AlphaCode \citep{li2022competition}: the \texttt{n@k}, which measures the fraction of problems the model can solve when allowed to generate $k$ solutions but only submit the best $n$ of them for evaluation. The results are shown in \Cref{tab:results}, and the ablation study results are shown in the bottom section of \Cref{tab:results}. We make the following observations.

\begin{figure}[h]
    \begin{center}
    \centerline{\includegraphics[width=\linewidth]{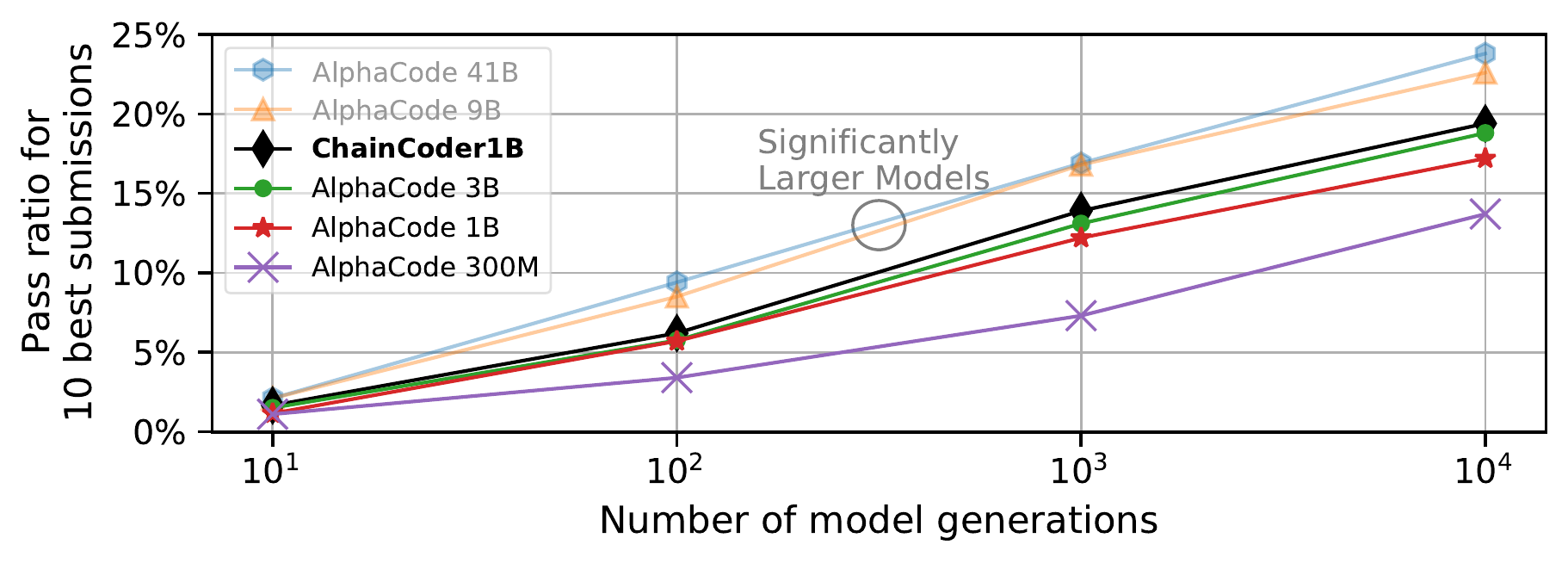}}
    \caption{Results on CodeContests dataset compared with AlphaCode ChainCoder with 1 billion parameters outperforms AlphaCode with similar sizes.}
    \label{fig:result-alphacode}
    \end{center}
\end{figure}

\renewcommand{\arraystretch}{1}
\begin{table*}[t]
    \centering
        \resizebox{0.9\textwidth}{!}{
            \begin{tabular}{l |c c | c c| c c| c c c c c}
                \toprule[1.5pt]
                & \multirow{2}{*}{Filtered From($k$)}  & \multirow{2}{*}{Attempts ($n$)} & \multicolumn{2}{c|}{Introductory} & \multicolumn{2}{c|}{Interview} & \multicolumn{2}{c}{Competition} \\
                & & & n@k & syntax-error-free & n@k & syntax-error-free & n@k & syntax-error-free \\
                \midrule
                GPT-3 few shot & N/A & 1 & 0.20\% & 31.0\% & 0.03\% & 42.0\% & 0.00\% & 40.0\% \\
                GPT-Neo 2.7B & N/A & 1 & 3.90\% & 87.9\% & 0.57\% & 87.9\% & 0.00\% & 85.0\% \\
                GPT-Neo 2.7B & N/A & 5 & 5.50\% & 97.4\% & 0.80\% & 96.0\% & 0.00\% & 95.4\% \\
                AlphaCode 1B & 1,000 & 5 & 14.36\% & N/A & 5.63\% & N/A & 4.58\% & N/A \\
                \midrule
            \textbf{ChainCoder 1B} &  1,000 & 5 & \textbf{17.52\%} & \textbf{100\%} & \textbf{7.36\%}  &  \textbf{100\%} & \textbf{5.48\%} & \textbf{100\%} \\

                \midrule
                ChainCoder I/O not aligned &  1,000 & 5 & 14.85\% & \textbf{100\%} & 5.90\%  &  \textbf{100\%} & 4.82\% & \textbf{100\%} \\
                ChainCoder with GPT code tokenizer &  1,000 & 5 & 14.80\% & \textbf{100\%} & 4.55\%  &  \textbf{100\%} & 3.46\% & \textbf{100\%} \\
ChainCoder no var-name replacement & 1,000 & 5 & 17.40\% & \textbf{100\%} & 7.25\% & \textbf{100\%} & 5.32\% & \textbf{100\%} \\
ChainCoder no destriptor during pre-train & 1,000 & 5 & 15.34\% & \textbf{100\%} & 6.29\% & \textbf{100\%} & 3.70\% & \textbf{100\%} \\

                \midrule
                ChainCoder tokenizer no warm-up &  1,000 & 5 & 16.10\% & \textbf{100\%} & 5.30\%  &  \textbf{100\%} & 4.22\% & \textbf{100\%} \\
ChainCoder tokenizer no $\bm{S}_1$ & 1,000 & 5 & 16.50\% & \textbf{100\%} & 5.80\% & \textbf{100\%} & 5.10\% & \textbf{100\%} \\
ChainCoder tokenizer no $\bm{S}_2$ & 1,000 & 5 & 16.90\% & \textbf{100\%} & 6.20\% & \textbf{100\%} & 5.22\% & \textbf{100\%} \\

                \midrule
                ChainCoder $\bm{S}_3$/$\bm{S}_4$ interleaved &  1,000 & 5 & 16.75\% & \textbf{100\%} & 6.80\%  &  \textbf{100\%} & 5.10\% & \textbf{100\%} \\


                \bottomrule[1.5pt]
            \end{tabular}
        }
    \caption{
    The results of APPS \cite{hendrycks2021measuring}, AlphaCode \cite{li2022competition} and ChainCoder on the APPS test set, fine-tuned on the APPS training set.
    }
    \label{tab:results}
\end{table*}

\textbf{ChainCoder Achieves Better Results.}
As seen in \Cref{tab:results} and \Cref{fig:result-alphacode}, ChainCoder reaches state-of-the-art performance. 
ChainCoder with one billion parameters outperforms the AlphaCode with similar size, and slightly outperforms AlphaCode with three billion parameters. With regard to the syntax-error-free rate, ChainCoder also stays competitive. As shown in \Cref{tab:results}, ChainCoder generated codes showcase impressive syntax pass rates on the APPS test set. This implies that the disentanglement of the layout frame and the accessory helps the model to better learn the syntax structure.

\textbf{ChainCoder Tokenizer Leads To Shorter Sequences Than Natural Language Tokenizer.}
Shorter sequence lengths intuitively points towards easier optimization and prediction for LLMs. We collect the sequence lengths of different tokenizers in \Cref{tab:stats}. The sequence length calculation of the ChainCoder tokenizer takes into account the components with complete information: \texttt{len($\bm{S}_3$) + len($\bm{S}_4$)}. We found that the ChainCoder's tokenizer encodes the program into shorter sequences, compared with the BERT tokenizer. For example, in the APPS dataset, the average sequence length is $78.404$ layout frame tokens +  $77.404$ accessory tokens ($155.808$ in total), whereas BERT tokenizer on average generates $176.312$ tokens. Though the sequence length will become longer if adding the coarse level $\bm{S}_1$ and $\bm{S}_2$ subsequences together, they do not add new prediction difficulty as they are subsets of $\bm{S}_3$.


\textbf{Syntax Aware Tokenization And I/O Aligned Sample Embedder Both Improve Model Performance.}
In this ablation study, we design ablation studies to justify the contributions of the design choices used in the tokenization step.
Specifically, we testify the coarse-to-fine syntax aware tokenization, the sample embedder with \iodata{} cross-sample alignment, the variable name replacement, the title summation based descriptor pre-training, and show the results in \Cref{tab:results}.
The \textit{ChainCoder with GPT code tokenizer} uses the natural language tokenizer for codes at the model output side, while the model input side uses the same tokenizer as the ChainCoder. The second variant, \textit{ChainCoder I/O not aligned} refers to the ablation study for the matrix-shaped aligned \iodata{} processed by the sample embedder. In this case, the \iodata{} is not aligned; instead, it is tokenized into a one-dimensional sequence with a natural language tokenizer. This sequence is then passed into the embedder as usual but with only one set of positional encodings.
The third variant \textit{ChainCoder no var-name replacement} does not apply the variable name replacement step. The fourth variant \textit{ChainCoder no destriptor during pre-train} do not feed input natural language summation to the model during the pre-training stage.

As can be seen from the results, \textit{ChainCoder with GPT code tokenizer} significantly underperforms the original ChainCoder. And when the \iodata{} alignment is dropped, the model also degrades. These results verify the contributions of the core model designs, i.e., the syntax-aware tokenization and the induced \iodata{} cross-sample alignment.
When no variable name replacement is applied, the performance is slightly worse, which suggests that in our case where the model learns to predict syntax format decoupled from the raw string text, learning the logical relationship between the variables are enough, and adding the semantic meaning of the variable names do not add performance to it. The performance for no descriptor is also unstable, which suggests the need for natural language from the pre-training stage.

\textbf{Coarse-To-Fine Generation Leads To Better Performance.}
In this ablation study, we replace our multi-step coarse-to-fine strategy with a single-shot final answer generation scheme. In other words, the ChainCoder only predicts steps $\bm{S}_3$ and $\bm{S}_4$, and omits $\bm{S}_1$ and/or $\bm{S}_2$. The result is represented as \textit{ChainCoder tokenizer no warm-up} (it has neither $\bm{S}_1$ nor $\bm{S}_2$), \textit{ChainCoder tokenizer no $\bm{S}_1$} and \textit{ChainCoder tokenizer no $\bm{S}_2$} in \Cref{tab:results}. As shown in the results, once the outline subsequence $\bm{S}_1$ and/or the core algorithm hint subsequence $\bm{S}_2$ are dropped, ChainCoder's performance takes a hit. This validates our assumption that warm-up with certain coarse information before generating the final results helps smooth the reasoning chain and improve the performance. The performance drop is especially prominent for more difficult benchmarks (the competition level benchmark), which further implies that difficult problems need more reasoning warm-up.

\textbf{Disentangling Layout Frame And Accessory Leads To Better Performance.}
In this case, we compare two settings but use the same tokenizer. The first setting is the original one, where $\bm{S}_1$, $\bm{S}_2$, $\bm{S}_3$, $\bm{S}_4$ are generated in four rounds, i.e., the prediction target is [$\cdots$, $\mathrm{<PAD>}$, $\bm{s}^3_1$, $\bm{s}^3_2$, $\cdots$, $\bm{s}^3_{N+1}$, $\mathrm{<PAD>}$, $\bm{s}^4_1$, $\bm{s}^4_2$, $\cdots$, $\bm{s}^4_{N}$, $\mathrm{<EOS>}$]. For the second setting shown in \textit{ChainCoder $\bm{S}_3$/$\bm{S}_4$ interleaved} in the bottom section of \Cref{tab:results}, the model is trained to generate a different sequence, where $\bm{S}_3$ and $\bm{S}_4$ interleaved together, just as originally shown in the syntax tree, i.e., the prediction target becomes [$\cdots$, $\mathrm{<PAD>}$, $\bm{s}^3_1$, $\bm{s}^4_1$, $\bm{s}^3_2$, $\bm{s}^4_2$, $\cdots$, $\bm{s}^3_{N}$, $\bm{s}^4_{N}$, $\bm{s}^3_{N+1}$, $\mathrm{<EOS>}$]. Comparing the results, we can see that separate generation performs better than interleaved version, validating the benefits of generating the layout frame and accessory separately.

\section{Conclusions}
In this work, we propose ChainCoder, a program synthesis language model that generates Python code progressively in multiple passes. The ChainCoder decouples the code snippet into the layout frame component and accessory component and uses them to construct a step-by-step, coarse-to-fine representation of the code as the prediction target. We leverage a novel transformer architecture to encode the syntactically aligned samples. Extensive evaluations showed that ChainCoder outperforms state-of-the-art code generation models, which verifies that generating related information before the final results as a warm-up helps to improve the language model in difficult code generation tasks.

\section*{Acknowledgement}

Z. Wang is in part supported by US Army Research Office Young Investigator Award W911NF2010240 and the NSF AI Institute for Foundations of Machine Learning (IFML).

\bibliography{refs}
\bibliographystyle{icml2023}

\newpage
\appendix
\onecolumn


\section{\iodata{} Augmentation Approach}
\label{sec:augmentation}

The model inputs include the natural language description and the \iodata{} of the given problem. Our natural language embedder uses a pre-trained BERT model, which has been trained on diverse texts. However, the sample embedder is only trained with \iodata{} sample tokens, and is only trained in the fine-tuning stage. 
To train the sample embedder with diverse texts, we carefully augment the \iodata{} and ensure that for each training instance in APPS and CodeContests, at least 100 program I/O pairs exist.
The data augmentation is only applied to the training set within the existing instances and does not cause data leakage.

We apply \textit{input distribution control} and \textit{output distribution control} to ensure the quality of the generated data. The input distribution control aims to estimate the underlying distribution of the cases provided officially and draw more data from it.
We implement a wide variety of data generators, and use a mixture of these generators to generate the inputs. We augment for both integer and string intputs, 
For example, if the program input is a list of integers, the values can be drawn from uniform distribution, quantized Gaussian distribution, almost sorted, all same value, almost having the same values with a few exceptions, or, the first element might be controlling the vector properties, such as the length of the vector, etc. We use different generators for these cases, and empirically assign probabilities to mix these generators.
The output distribution control is only applied to bool program outputs. For each generated input sample, we use the existing code in the dataset to run and get the outputs. If one label (true or false) is significantly more than the other in the augmented data, then more is dropped to make sure the output labels are balanced.

\section{Token Compositions}

\begin{figure}[!h]
\begin{center}
\centerline{\includegraphics[trim={2cm 4.5cm 2cm 2.5cm},clip,width=\columnwidth]{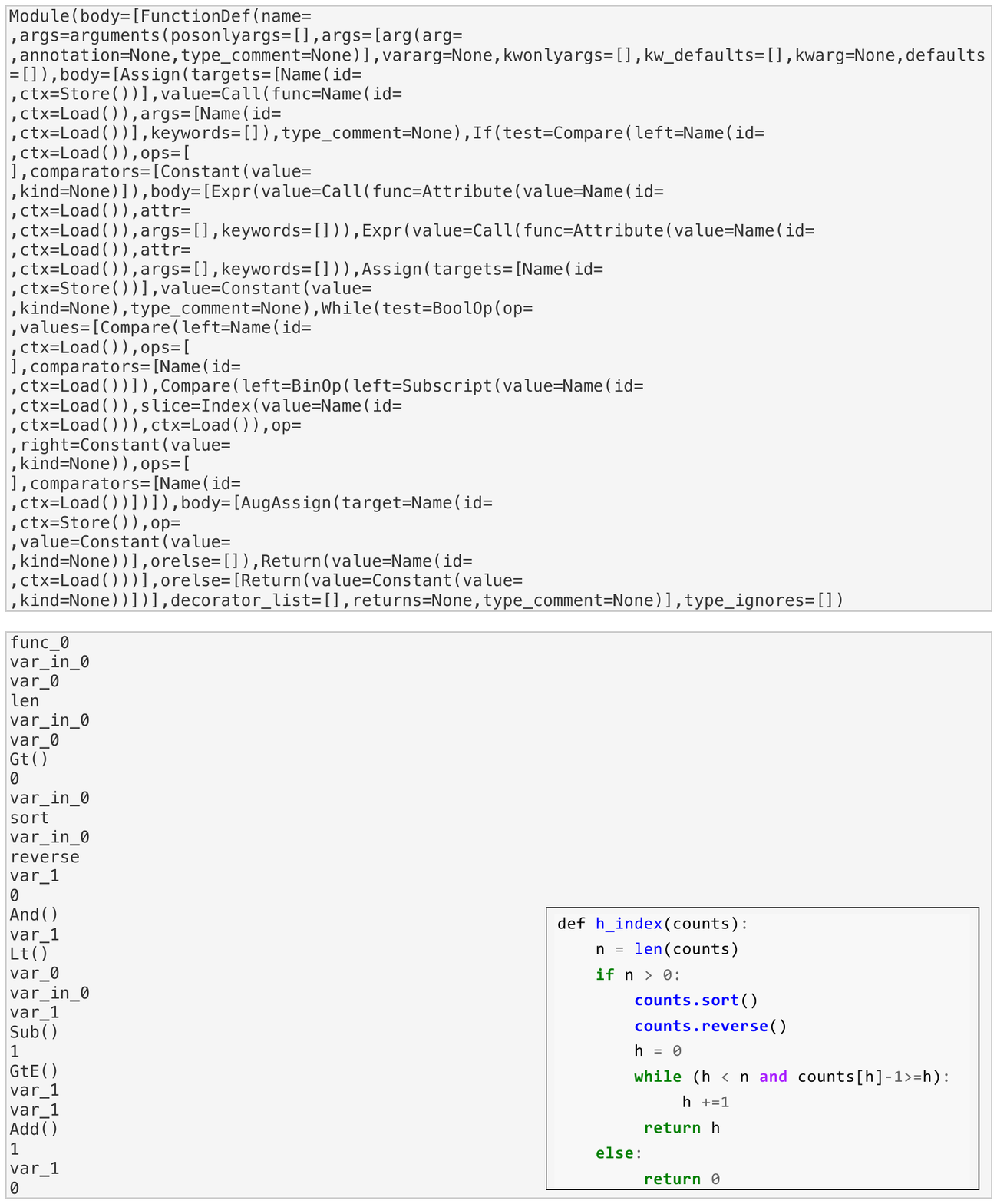}}
\caption{Token strings demonstration: the strings in the above and below figures are $\bm{S}_{3}$ and $\bm{S}_{4}$ subsequences for the code snipped in the bottom right, respectively.} 
\vspace{-1em}
\label{fig:demo1}
\end{center}
\end{figure}

The encoded token strings for the  $\bm{S}_{3}$ and $\bm{S}_{4}$ subsequences are visualized in \Cref{fig:demo1}.

The ChainCoder leverages a novel tokenizer based on AST instead of natural language. More specifically, the code expert tokenizer is used in the sample embedder to encode the program \iodata{}, and used in the model output side to encode the program syntax tree.

The vocabulary comprises of two parts: the tokens for the layout frame subsequence, which are grouped branching nodes in the syntax tree, and the tokens for the accessory subsequence, which are the leaf nodes. The layout frame part is collected by sweeping the dataset, as shown in \Cref{fig:vocab-dist}, and the accessory part is pre-computed and stored. The composition of the accessory part is shown in \Cref{table:content}.

\begin{table}[b]
    \resizebox{\textwidth}{!}{
    \centering
    \begin{tabular}{l|l|l}
        \toprule[1.5pt]
        \textbf{Type}                   & \textbf{Definition}                                                & \textbf{Example}                                \\ \midrule
        Names                  & The names of variables, functions, classes, operators, etc., replaced from name pool & ``var\_1'', ``func\_1'', ``class\_1'', ... \\
        Build in python vocabs & About 600 common python's built-in functions and keywords & ``\_\_name\_\_'', ``\_\_package\_\_'', ... \\ 
        Digit                  & Int representation of single digit number                 & 0,1,2,3,4,5,6,7,8,9                    \\ 
        ASCII                  & Character encoding                                        & ``a'', ``!'', ``='', ``0'', ``{[}SPACE{]}'',...  \\ 
        Common float           & Float that appears most common while training             & 0.1, 0.0001, 0.5,  0.2, ...          \\ 
        \bottomrule[1.5pt]
    \end{tabular}
    }
    \caption{Types of accessory tokens}
    \label{table:content}
\end{table}




\section{Visualizations}

On the input side, for each given instance, multiple \iodata{} could be provided as part of the problem specifications. As the \iodata{} sequence is python-interpretable, it is also processed by the AST based tokenizer. Thanks to the syntax tree structure, they can be aligned across samples based on their syntax roles. The \iodata{} cross-sample alignment is shown in \Cref{tab:glue}. 

\begin{figure*}[b]
    \centering
    \includegraphics[width=0.9\textwidth]{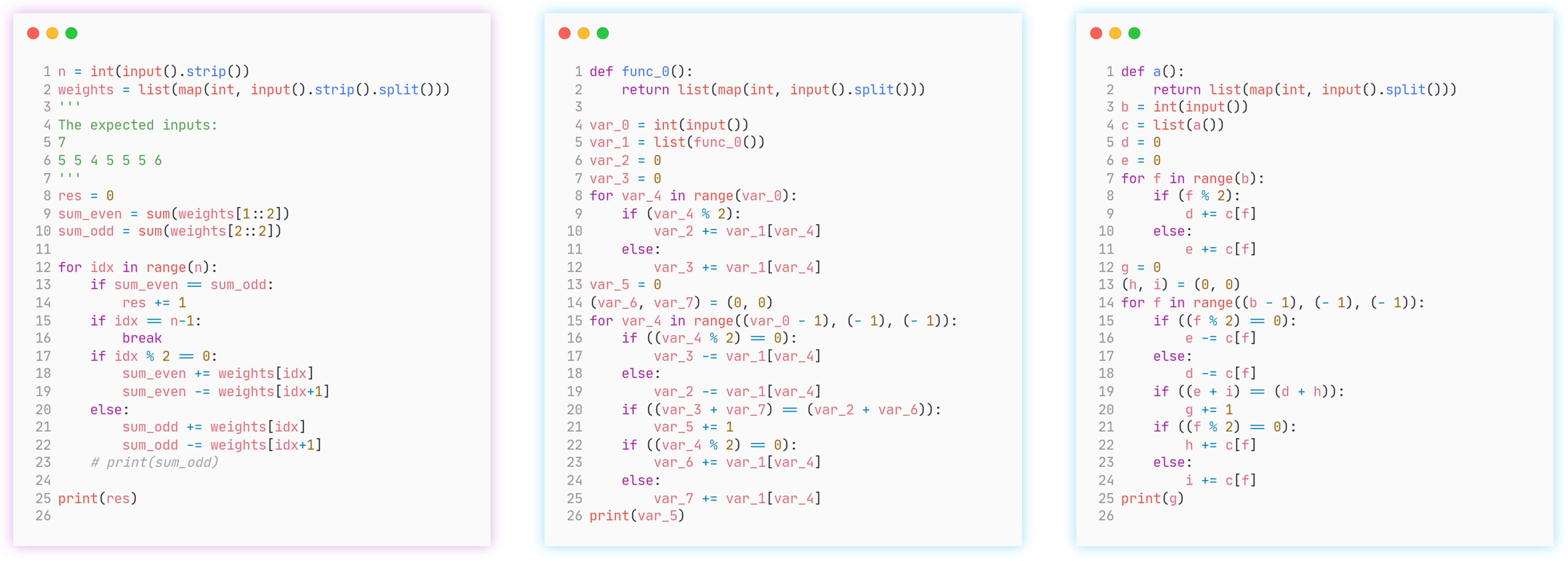}
    \caption{The example outputs of APPS model (GPT-Neo 2.7B, left) and ChainCoder (middle and right, with two different post-naming rules), over the same problem.}
    \label{fig:demo outputs}
\end{figure*}

In addition, the examples of generated code are shown in \Cref{fig:demo outputs}, where we pick the is-palindrome problem and compare the code generated by GPT Neo and ChainCoder.

\renewcommand{\arraystretch}{1.25}
\begin{table*}[t]
\centering
\resizebox{0.95\textwidth}{!}{
\begin{tabular}{l|l|l|l}
\toprule[1.5pt]
Sample                    & Input                    & Content                                & Syntax                                                                           \\ \hline
\multirow{9}{*}{sample 1} & \multirow{5}{*}{Input 1} & $<$align-padding$>$ & $<$align-padding$>$                                             \\ \cline{3-4} 
                          &                          & $<$align-padding$>$ & $<$align-padding$>$                                             \\ \cline{3-4} 
                          &                          & 0                                      & Module(body={[}Expr(value=List(elts={[}List(elts={[}List(elts={[}Constant(value= \\ \cline{3-4} 
                          &                          & 2                                      & kind=None),Constant(value=                                                       \\ \cline{3-4} 
                          &                          & 3                                      & kind=None),Constant(value=                                                       \\ \cline{2-4} 
                          & \multirow{4}{*}{Input 2} & $<$align-padding$>$ & $<$align-padding$>$                                             \\ \cline{3-4} 
                          &                          & $<$align-padding$>$ & $<$align-padding$>$                                             \\ \cline{3-4} 
                          &                          & 0                                      & kind=None){]},ctx=Load()),Constant(value=                                        \\ \cline{3-4} 
                          &                          & $<$wait-padding$>$             & kind=None){]},ctx=Load()){]},ctx=Load())){]},type\_ignores={[}{]})              \\ \hline
\multirow{9}{*}{sample 2} & \multirow{5}{*}{Input 1} & 0                                      & Module(body={[}Expr(value=List(elts={[}List(elts={[}List(elts={[}Constant(value= \\ \cline{3-4} 
                          &                          & 2                                      & kind=None),Constant(value=                                                       \\ \cline{3-4} 
                          &                          & 3                                      & kind=None),Constant(value=                                                       \\ \cline{3-4} 
                          &                          & 5                                      & kind=None),Constant(value=                                                       \\ \cline{3-4} 
                          &                          & 1                                      & kind=None),Constant(value=                                                       \\ \cline{2-4} 
                          & \multirow{4}{*}{Input 2} & $<$align-padding$>$ & $<$align-padding$>$                                             \\ \cline{3-4} 
                          &                          & $<$align-padding$>$ & $<$align-padding$>$                                             \\ \cline{3-4} 
                          &                          & 2                                      & kind=None){]},ctx=Load()),Constant(value=                                        \\ \cline{3-4} 
                          &                          & $<$wait-padding$>$             & kind=None){]},ctx=Load()){]},ctx=Load())){]},type\_ignores={[}{]})              \\ \hline
\multirow{9}{*}{sample 3} & \multirow{5}{*}{Input 1} & $<$align-padding$>$ & $<$align-padding$>$                                             \\ \cline{3-4} 
                          &                          & 4                                      & Module(body={[}Expr(value=List(elts={[}List(elts={[}List(elts={[}Constant(value= \\ \cline{3-4} 
                          &                          & 5                                      & kind=None),Constant(value=                                                       \\ \cline{3-4} 
                          &                          & 6                                      & kind=None),Constant(value=                                                       \\ \cline{3-4} 
                          &                          & 7                                      & kind=None),Constant(value=                                                       \\ \cline{2-4} 
                          & \multirow{4}{*}{Input 2} & 5                                      & kind=None){]},ctx=Load()),Constant(value=                                       \\ \cline{3-4} 
                          &                          & 3                                      & $<$wait-padding$>$                                              \\ \cline{3-4} 
                          &                          & 2                                      & $<$wait-padding$>$                                              \\ \cline{3-4} 
                          &                          & $<$wait-padding$>$            & kind=None){]},ctx=Load()){]},ctx=Load())){]},type\_ignores={[}{]})'              \\ \bottomrule[1.5pt]
\end{tabular}}
\caption{One instance with three exsamples of I/O data. 
\texttt{io1 = [[[0,2,3],0], [12, 'abcd']]}, 
\texttt{io2 = [[[0,2,3,5,1], 2], [43, 'm']]}, 
\texttt{io3 = [[[4,5,6,7], 532], [9908, 'ss']]}. }
\label{tab:glue}
\end{table*}

\begin{figure}[t]
    \centering
    \subfigure[Full solution code]{
        \includegraphics[width=0.22\columnwidth]{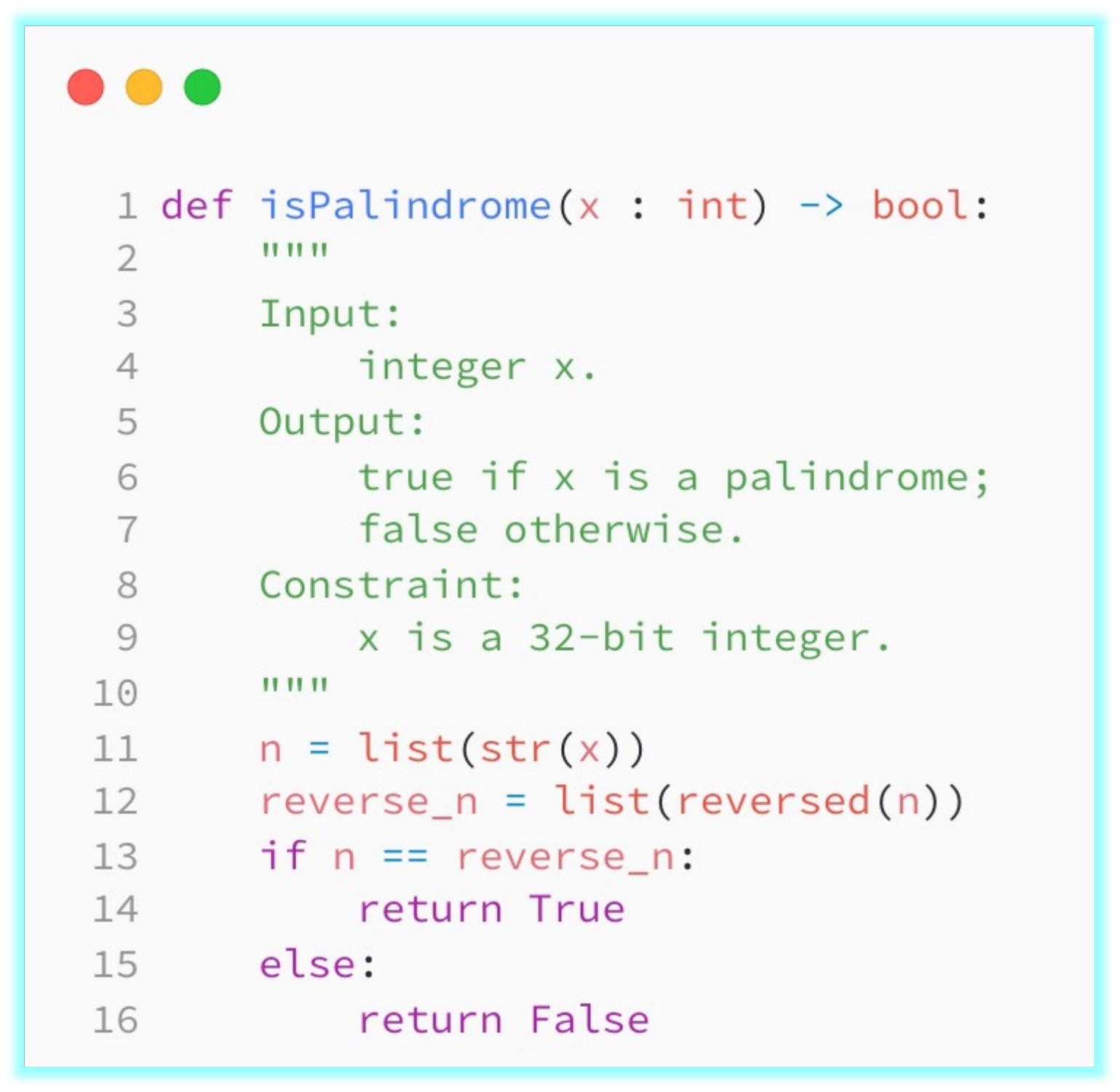}
    }    
    \subfigure[Syntax tree of one line]{
        \includegraphics[width=0.12\columnwidth]{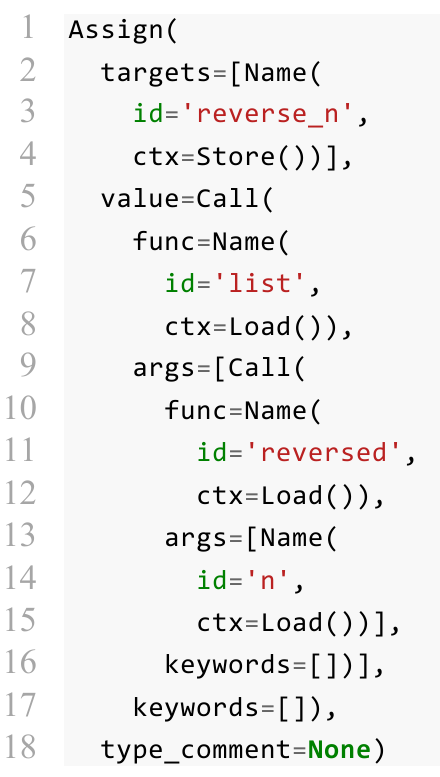}
    }
    \subfigure[Illustration of natural language and ChainCoder tokenizers]{
        \includegraphics[width=0.6\columnwidth]{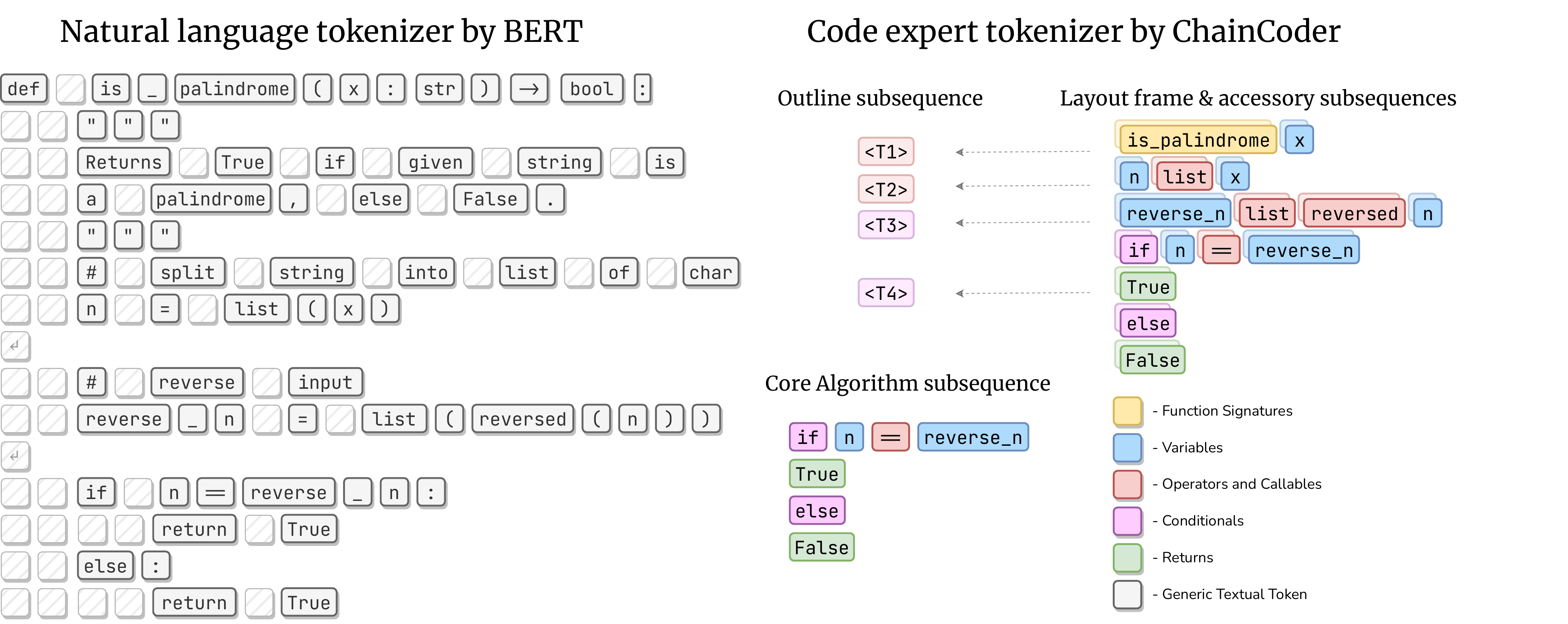}
    }

    \caption{One Python solution to the is-palindrome example problem (left), and a subtree of the AST parsed syntax tree, from the sentence \texttt{reverse\_n = list(reversed(n))} (right). The full tree is in \Cref{fig:LARGEAST} in the Appendix with 106 lines in total.}
    \label{fig:SMALLAST}
\end{figure}

\begin{figure*}[t]
    \centerline{\includegraphics[trim={1.7cm 2cm 2.8cm 3cm},clip,width=0.9\columnwidth]{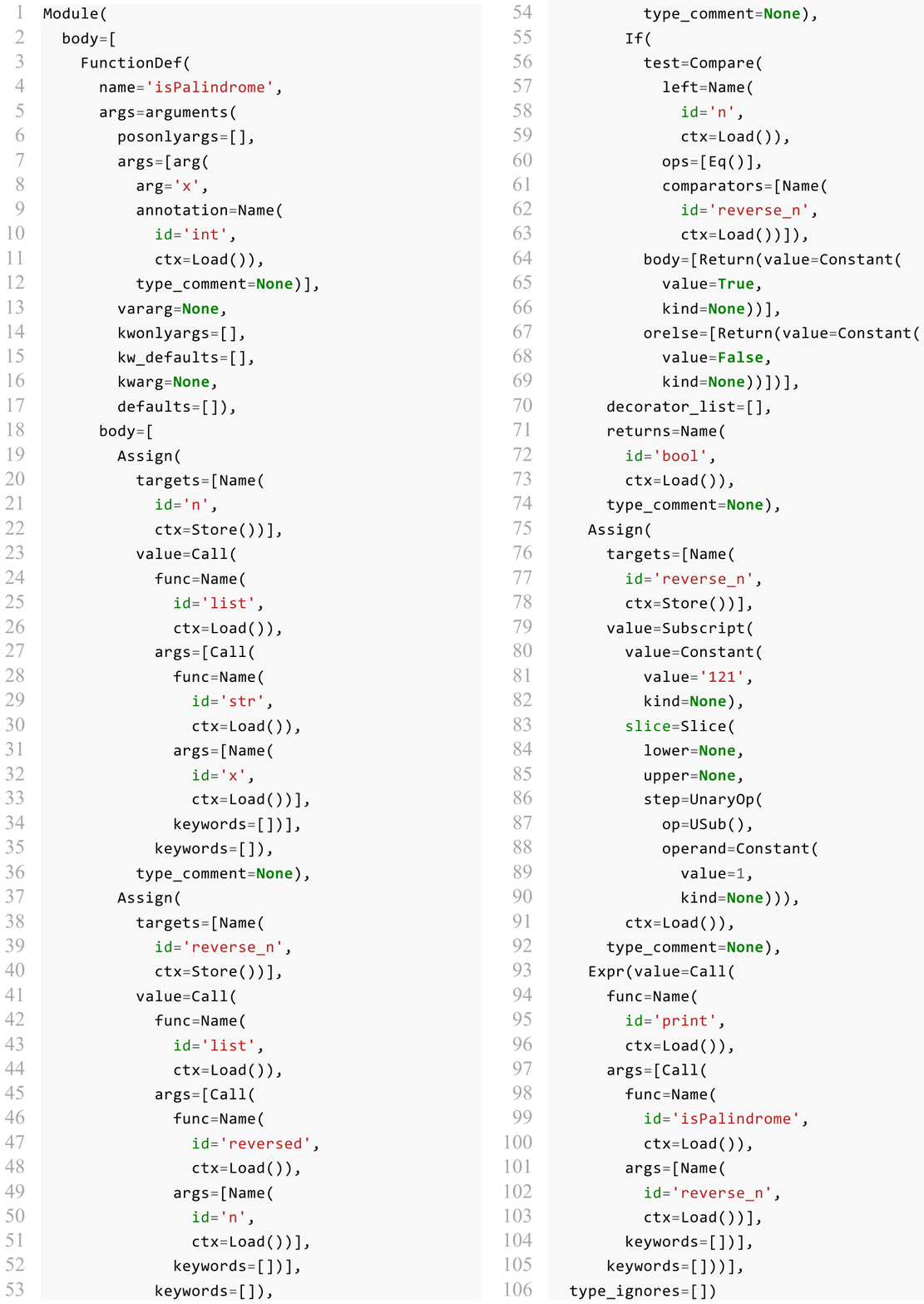}}
    \caption{Complete format of our AST tree, in the example of \texttt{isPalindrome} case. The corresponding code is in \Cref{fig:SMALLAST}.}
    \label{fig:LARGEAST}
\end{figure*}


\end{document}